\documentclass[runningheads]{llncs}
\usepackage[T1]{fontenc}
\usepackage{graphicx}
\usepackage{hyperref}
\usepackage{color}

\usepackage{url}
\usepackage{booktabs}
\usepackage{multirow}
\usepackage{makecell}
\usepackage{pifont}
\usepackage[inline]{enumitem}
\usepackage{siunitx}
\usepackage[subtle, mathspacing=normal]{savetrees}
\newcommand{\vspaceSecBefore}[0]{}
\newcommand{\vspaceSecAfter}[0]{}

\usepackage{tablefootnote}
\usepackage{caption}
\usepackage{amsmath}

\begin{document}
%
\title{On LLM-Assisted Generation of\\ Smart Contracts from Business Processes
}
\titlerunning{LLMs for Smart Contract Generation from Business Processes}
%
%
\authorrunning{Stiehle et al.}
%
\author{Fabian Stiehle\inst{1} \and Hans Weytjens\inst{1} \and Ingo Weber\inst{1,2}}
%
%
%
\institute{
Technical University of Munich, School of CIT, Germany, \email{first.last@tum.de} \and
    Fraunhofer Gesellschaft, Munich, Germany
}
%
\maketitle              
\begin{abstract}
Large language models (LLMs) have changed the reality of how software is produced. Within the wider software engineering community, among many other purposes, they are explored for code generation use cases from different types of input. 
In this work, we present an exploratory study to investigate the use of LLMs for generating smart contract code from business process descriptions, an idea that has emerged in recent literature to overcome the limitations of traditional rule-based code generation approaches.
%
However, current LLM-based work evaluates generated code on small samples, relying on manual inspection, or testing whether code compiles but ignoring correct execution.
With this work, we introduce an automated evaluation framework and provide empirical data from larger data sets of process models. We test LLMs of different types and sizes in their capabilities of achieving important properties of process execution, including enforcing process flow, resource allocation, and data-based conditions.
Our results show that LLM performance falls short of the perfect reliability required for smart contract development. We suggest future work to
explore responsible LLM integrations in existing tools for code generation to ensure more reliable output. Our benchmarking framework can serve as a foundation for developing and evaluating such integrations.
\end{abstract}
\vspaceSecBefore
\section{Introduction}
\vspaceSecAfter
In many fields, LLMs are envisioned to assist in a wide variety of tasks, where so far the training of general machine learning models has been challenging due to the scarcity of 
specialized data and the large computational effort required (c.f., \cite{belzner2023large,burgueno2025automation,vidgof2023large}).

In the wider field of software engineering, LLMs are anticipated to support all phases of the software engineering process~\cite{belzner2023large}. 
For code generation, LLMs have arguably made the biggest practical impact so far, testified by the integration of commercial tools like \textit{Github’s Copilot}---which builds on seminal research on LLMs trained on code~\cite{chen2021evaluating}---into popular development environments like \textit{Visual Studio Code}.

Similarly, within the field of business process management (BPM), Vidgof et al.~\cite{vidgof2023large} call to evaluate the combination of LLMs with existing BPM technologies. First results evaluating LLMs on BPM tasks show the large promise, since they perform comparable to or better than existing BPM tools~\cite{grohs2023large}.

Blockchain-based business process execution relies on a model-driven paradigm, where process descriptions are transformed into executable artefacts based on rule-based transformation tools~\cite{stiehle2022blockchain}. These tools, however, exhibit various limitations such as in their flexibility, e.g., in terms of supported process modelling constructs, their supported output targets, or their support of blockchain-specific features. Considering the wider field of model-driven engineering, LLMs are similarly prospected to have the potential to drive automation~\cite{burgueno2025automation}. This hope can draw on seminal results from the related field of code-to-code translation (transpilers), where LLM-based approaches were found to outperform traditional rule-based approaches~\cite{roziere2020unsupervised,roziere2021leveraging}.

While many positive visions exist, and early results on leveraging LLMs to assist in code generation are impressive~\cite{brynjolfsson2023copilot}, significant challenges remain—extending even beyond the well-known hallucination issue (or more precisely: confabulation~\cite{Smith:2023:Confabulation}). For instance, GitHub Copilot can introduce numerous security vulnerabilities into generated code~\cite{pearce2025asleep}. LLM outputs are inherently non-deterministic~\cite{ouyang2025empirical} , making them unreliable for consistent behaviour. Meanwhile, Huang et al. \cite{huang2024bias} show that generated code may reproduce ethically concerning biases, such as gender-related ones.

Furthermore, proprietary models raise concerns about confidentiality, privacy, and autonomy (c.f.~\cite{belzner2023large,haase2025sustainability}). These AI models are often deployed on large, centralised platforms provided by hyperscalers like \textit{AWS}, \textit{Azure}, or \textit{GCP}. Open-source models running on these platforms face the same security concerns.
Relying on central deployments may not be a good fit for blockchain-based processes, where decentralisation is a goal~\cite{2024-Stiehle-BPMWS}.

Thus, it is paramount to systematically evaluate LLMs' usefulness and fit of properties for a given task and weigh benefits and drawbacks. Research is called to identify the scenarios where LLMs add true value (c.f. \cite{belzner2023large,vidgof2023large}). However, evaluation poses a significant challenge, as it often requires manual human investigation and large volumes of labelled or parallel data~\cite{chang2024survey}.

Research on using LLMs for smart contract generation is in its infancy. Existing work aims, among other things, to reduce the required specialised programming skills for contract development. However, most of this work focuses on reporting the syntactical correctness of generated code (compilability), not its functional correctness~\cite{alam2025solgen,luo2025guiding,napoli2024leveraging}. In other cases, correctness is assessed through manual inspection of small samples~\cite{chatterjee2025efficacy,gao2025bpmn}.
In the case of blockchain-based process execution, early work explores the promise of deriving code from process descriptions, but draws conclusions from a singular case study~\cite{gao2025bpmn}.
 Current research furthermore does not make data or code openly accessible.

In this work, we present an exploratory study investigating the use of LLMs to generate smart contract code from business process descriptions. We think it paramount to ground such research in open empirical evidence from larger datasets.
Beyond standards of open science, the availability of open repeatable benchmarks and data is especially important in the present case, as  LLMs capabilities develop in a fast pace and their output is unpredictable.
With this work, we introduce an automated evaluation framework for generating smart contract code from process models. We provide empirical data from larger data sets of process models (165 models filtered and sampled from the collection of SAP-SAM~\cite{sap-sam} models). 
We test seven LLMs of different types and sizes in their capabilities of achieving central properties of blockchain-based process enactment (c.f.~\cite{stiehle2022blockchain}): enforcing process flow, case data-based conditions, resource allocation, and efficiency.

In general, our results indicate good performance of some models, with those achieving F1 scores of 0.8 or more.
Due to the stochastic nature of LLMs, output remains imperfect and unreliable. While such performance may be acceptable in other contexts, blockchain is an unforgiving environment---on public blockchains developers should assume that any weakness will be exploited.
As such, we believe this is a fundamental issue of the chosen approach to have LLMs generate smart contract code, which cannot be overcome by LLMs based on the current architectures.
We discuss this point further in the paper, including an outlook on roles LLMs could fulfil productively in the generation of smart contracts from process models. 

The remainder of the paper is structured as follows. A background on LLMs and relevant terminology is provided in the next section; note that we assume familiarity with blockchain and process enactment on it. 
Subsequently related work is discussed in Section~\ref{sec:relw}, before we present the benchmarking framework in Section~\ref{sec:framework}.
Based on it, we conduct experiments that we report on in Section~\ref{sec:eval} and discuss in Section~\ref{sec:discussion}, before Section~\ref{sec:concl} concludes.

\vspaceSecBefore
\section{Background}\label{sec:bg}
\vspaceSecAfter
In this section, we give relevant background on the AI models, their attributes, and related concepts which we use in the body of the paper.
Large Language Models (LLMs) are transformer-based neural network systems~\cite{vaswani2017attention}. LLMs are a specific class of Foundation Models, a class of machine learning models trained on extensive data sets, not for one specific purpose but as a basis for many possible applications~\cite{AI-Engineering-Book-2025}.
In the case of LLMs, training ingests very large textual corpora, comprising not only natural language, but also programming code (such as Solidity for blockchain smart contracts), and formal representations (such as BPMN for modeling business processes). Due to their generality and the training input, LLMs can perform complex tasks beyond language understanding and generation, including analyzing and creating business process models~\cite{horner2025towards} and smart contracts~\cite{de2025llm}.
Many of the latest LLMs support other modalities (such as images, video, or audio), hence they are sometimes referred to as large multi-modal models (LMMs). However, for this paper, the distinction between LLMs and LMMs is of no importance, and hence for the sake of clear communication, we follow the current common practice and refer to them as LLMs.

In many usage scenarios, LLMs deliver high performance out of the box, eliminating the need for fine-tuning or training from scratch with large supervised datasets or significant compute resources. In \textit{zero-shot prompting}, models complete tasks based solely on instructions without prior examples, whereas \textit{few-shot prompting} involves providing a handful of illustrative examples directly within the input~\cite{dong2024surveyincontextlearning}.

\textit{Tokens} are the units of input and output for LLMs. Tokens can be whole words, subwords (parts of words), individual characters, punctuation, or special characters~\cite{ali2024tokenizer}. The total number of tokens---both in the input query and the generated output---directly influences the computational load of running these models. 
Models that perform complex \textit{reasoning} (so called reasoning models) typically generate a plan to answer a query, execute the steps in the plan, and possibly check their work; hence they require many more tokens than regular LLMs. 
Increasing the model size---measured by the number of parameters, the numeric values representing the strength of connections between ``neurons''---generally improves performance but also raises computational demands, leading to greater energy consumption~\cite{odonnell2025energy}. Balancing model size, token usage, and capability is therefore essential for developing efficient and sustainable LLM-based applications.

LLMs are available in both proprietary and open-source forms~\cite{wong2024comparative}. Proprietary models, such as \textit{OpenAI's GPT} and \textit{Anthropic's Claude} model families, are typically accessed via APIs hosted by third parties. In comparison to open-source models, they often deliver superior performance but introduce risks such as data exposure, dependency on external providers, and limited transparency. In contrast, open-source models enable self-hosting and on-premise deployment---an attractive option in blockchain contexts where data sovereignty, trust minimization, and operational independence are essential. 

\textit{Temperature} is a setting for LLMs that controls the randomness of generated text; lower temperatures make outputs more predictable and focused (by selecting the most probable tokens), while higher temperatures encourage more creative and varied responses (by selecting less probable tokens).

\vspaceSecBefore
\section{Related Work}\label{sec:relw}
\vspaceSecAfter
Within the field of BPM, Vidgof et al.~\cite{vidgof2023large} outline the opportunities and challenges of integrating LLM-based tools within the BPM lifecycle. 
Recent work explores an increasingly wide array of BPM applications (see e.g., Pfeiffer et al.~\cite{pfeiffer2025theorypracticerealworlduse}, which explore four real-world use cases that demonstrate the use of LLMs across modelling, prediction and automation). Within predictive process monitoring, LLMs are explored for their capability to predict future states of processes (e.g., Pasquadibisceglie et al.~\cite{10680620}). In prescriptive process monitoring, LLMs are explored to enhance recommendations with LLM-generated explanations~\cite{10.1007/978-3-031-70396-6_23}. 

Orthogonal to our work is the question whether LLMs can assist in deriving accurate models from natural language process descriptions (e.g., Hörner et al~\cite{horner2025towards}); for an overview see Klievtsova et al.~\cite{klievtsova2023conversational}.
Closer to our work, Monti et al.~\cite{monti2024nl2processops} propose an LLM-driven pipeline to extract executable scripts (for deployment in a process execution engine) from natural language process descriptions. For the evaluation of their code generation, they use 10 cases and compare the LLM output to a manually implemented script. Additionally, they conducted a human evaluation, assessing the quality of the produced code. A similar study to ours was conducted by Berti et al.~\cite{berti2024pm}, which presents a benchmark on the performance of LLMs for different Process Mining tasks, the open-ended nature of these tasks requires a \textit{LLM judge} evaluation approach, where LLMs assess the output of other LLMs.

Research on LLM-assisted smart contract generation is in its infancy. Existing results are limited to reporting the syntactical correctness of generated code (compilability) and the automated detection of known vulnerabilities using existing tools~\cite{alam2025solgen,napoli2024leveraging,luo2025guiding}. A relevant result to our investigation is provided by Luo et al.~\cite{luo2025guiding}; their result suggests that augmenting the generation process by a formalised model improves results.
Some works go beyond syntactical correctness by manual inspection of code~\cite{chatterjee2025efficacy,gao2025bpmn}. However, this is limited to a few simplistic cases, as it is highly labour intensive.
Karanjai et al.~\cite{karanjai2023smarter} extract solidity functions from GitHub repositories and use extracted code comments as prompts; to test the LLM output, they rely on corresponding unit tests present in the mined repository.

None of the aforementioned studies on smart contract generation make their data  or evaluation frameworks available.

Our work lies at the intersection of previously outlined fields. To the best of our knowledge, the only directly related work is the recent case study of Gao et al.~\cite{gao2025bpmn}. They present an approach, based on few-shot prompting, capable of generating a smart contract from a Business Process Modeling Language (BPML) specification (a superset of BPEL), derived from a collaboration diagram. However, their evaluation is limited to one process, which includes start, end, and message events, tasks, and two XOR splits. They do not make their code or data available.

To the best of our knowledge, we are the first to present an open source benchmarking framework to automate the assessment of code generation from process models. We provide an instantiation of it based on the large SAP-SAM data set~\cite{sap-sam}, one-shot and two-shot prompts, and a Ethereum virtual machine (EVM) environment. Using it, we evaluate different proprietary and open source LLMs on $165$ cases each. All our data is available and our tests can be repeated (see Footnote~\ref{fn:repo}).


\vspaceSecBefore
\section{Benchmarking Framework}\label{sec:framework}
\vspaceSecAfter
To assess the capabilities of LLMs for smart contract generation, we designed a configurable benchmarking framework and make it openly available along with all relevant input and output data.\footnote{\label{fn:repo}\url{https://github.com/fstiehle/bpmn-sol-llm-benchmark}. An archived version is available at: \url{https://doi.org/10.5281/zenodo.16616694}.} An open framework facilitates repeatability and can be used to judge the capabilities of future model evolutions. We envision our framework to also serve as a foundation for evaluation in future related research (as outlined in Section~\ref{sec:discussion}). In contrast to previous work, we want to facilitate automated evaluation from large data sets with optimal coverage. An established method to benchmark the correctness of a blockchain-based business process is to replay all possible conforming traces\footnote{For the remainder of the paper, we use common terminology. We loosely denote an event log as a set of events, where each event is associated with a case identifier that groups it into a case. A trace is the ordered sequence of events (activities) that occurred for a specific case. Each event represents a task (activity) in the model. A trace can be said to be in conformance with a process model if it represents a valid execution path through that model.} (which the smart contract has to accept) and replay a set of non-conforming traces (which the smart contract has to reject)~\cite{stiehle2022blockchain}. Our approach makes use of this method. From a given process model, it generates conforming traces (not always exhaustively, but up to a configurable threshold, as parallelism and loops can result in an intractable search space) and non-conforming traces. All traces are then replayed against the generated smart contracts.
\subsection{Architecture Overview}
\begin{figure}[tb]
    \centering
    \includegraphics[width=.95\linewidth]{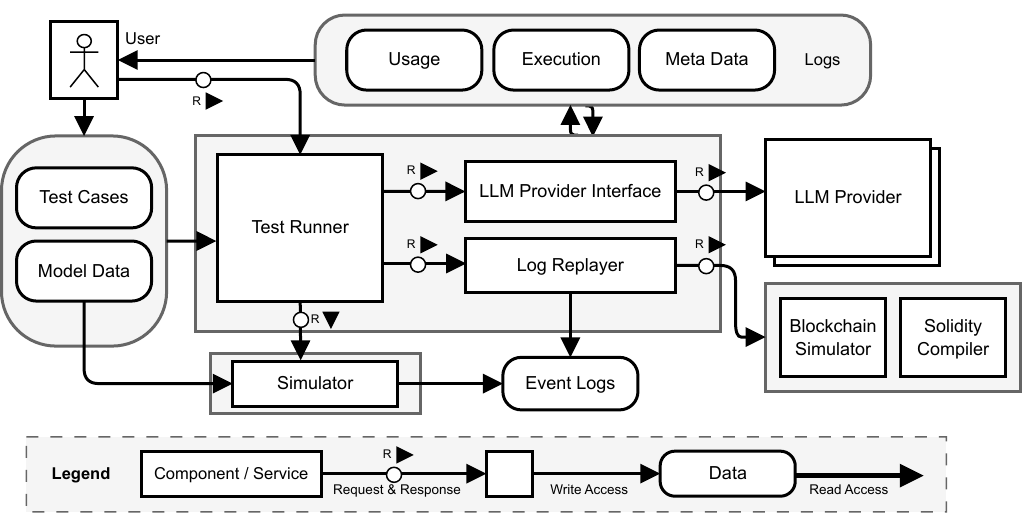}
    \caption{Main components, services and data of the benchmarking framework architecture as FMC block diagram.}
    \label{fig:architecture}
\end{figure}
Figure~\ref{fig:architecture} depicts an overview of our architecture. As input, different test cases can be configured. Mainly, a test case is a tuple of: (i) the LLM to benchmark, (ii) the prompt to use, (iii) the process model data set to use.

Interaction with the framework is achieved through a \textit{test runner}. 
Based on user interaction, the runner coordinates the benchmark execution with the other internal and external components. From the model data (process models), a \textit{simulator} component generates conforming and non-conforming traces. This simulator component is external to the framework, so it can be swapped based on different model data inputs.
The simulator is responsible for generating conforming and non-conforming event logs from the input process models. To generate a non-conforming trace, a conforming trace is randomly chosen and manipulated. The simulator also generates an encoding that maps the events and participants to how they should be represented in the smart contract (\verb|taskID|s and \verb|participantID|s, the latter associated with a blockchain address).

This encoding is embedded into the prompt, along with the model data, by the \textit{LLM provider interface}, which calls the external LLM provider as configured in the test cases. The received output is stored in a usage log. For any interaction with an LLM provider, the usage log stores (among other things):
a timestamp, the configured test case, the full input and the full output. This allows to reconstruct and rerun a benchmark at a later time. The usage log also records usage statistics (tokens used, price, etc.) as reported by the provider.

From the usage log output, the \textit{log replayer} extracts the smart contract code, compiles and deploys it to an external blockchain environment, and uses the encodings and event logs to perform a benchmark on the deployed contract. The replayer stores the results in execution logs. 
The framework also extracts metadata on the used process model (e.g, modelling constructs per model, etc.) to aid in the evaluation of the benchmark.

\subsection{Instantiation}
In the current instantiation of our framework, we support BPMN 2.0 Choreographies. This is a purely practical implementation choice, given that there is no consensus on the best fitting modelling paradigm for blockchain-based execution (c.f.~\cite{stiehle2022blockchain}) and given our familiarity with a suitable tool.
We instantiate our framework for an Ethereum virtual machine (EVM) blockchain environment, the most widely employed environment~\cite{stiehle2022blockchain}.
We discuss key design decisions in the instantiation of our architecture.

\paragraph{Simulator.}
We extend the open source tool \textit{Chorpiler}, first introduced in~\cite{stiehle2023processchannels} with simulation capabilities. Chorpiler transforms BPMN Choreographies to smart contracts, generates non-conforming traces from conforming traces, and also generates machine-readable encodings on how to interact with a contract. Chorpiler parses a given Choreography into an interaction net, a special type of labelled Petri net (see~\cite{decker2007local}), suitable to represent choreographies. We make use of this intermediate presentation to generate conforming traces. Here, we adopt the implementation of \textit{pm4py}~\cite{pm4py}, a popular Python library for process mining, which includes a \textit{playout} functionality to generate event log traces from Petri nets. The implementation tries to discover new conforming traces with each pass until a threshold of passes is reached.

As we also want to benchmark data-based exclusive gateways (XOR), we had to extend the playout functionality to generate appropriate data manipulation events. To do so, for each outgoing flow (other than the default flow), we generate a boolean decision. During trace generation, once our algorithm encounters a decision transition, it inserts a corresponding data event in the trace.\footnote{Chorpiler implements transactional logic as in~\cite{ladleifModelingEnforcingBlockchainBased2019}; the smart contract makes as much progress as possible after a task is executed; i.e., data-based decisions are made autonomously once the gateway is enabled, and the required data must be already set by then. Thus, data events are inserted preceding any event that leads to a given gateway.} 


\paragraph{Test Runner, Replayer and LLM Provider.}
To provide the replayer with a blockchain environment, we use \textit{hardhat}, a popular Ethereum development framework that allows testing, deployment, and debugging of smart contracts in a locally simulated EVM environment.\footnote{\url{https://hardhat.org}, accessed 2025-06-12} As LLM provider, we use \textit{OpenRouter}, a platform that provides a unified API across multiple language model providers. This simplifies our integration and gives access to a broad range of state-of-the-art models.\footnote{\url{https://openrouter.ai}, accessed 2025-06-12} 
We use a \textit{Node.js} environment; the test runner is implemented using \textit{Mocha}, a JavaScript test framework.\footnote{\url{https://www.npmjs.com}, \url{https://mochajs.org}, both accessed 2025-06-12}

%
\vspaceSecBefore
\section{Experiment}\label{sec:eval}
\vspaceSecAfter

We use the instantiation of our framework to conduct a large scale benchmarking experiment. 
Our process model data is based on the \textit{SAP Signavio Academic Models Dataset} (SAP-SAM), which was initially introduced in~\cite{sap-sam}. The dataset contains different model types created through the \textit{Academic Initiative} platform from 2011 to 2021.\footnote{\url{https://academic.signavio.com}, accessed 2025-06-12} Thus, the dataset primarily contains process models created by students, researchers, and teaching staff. Still, for BPMN specifically, the dataset's properties (distribution of modelling constructs) are in line with previous research assessing the usage of modelling constructs from diverse sources~\cite{Muehlen2013}.
The collection includes \num{4096} BPMN 2.0 choreography models, to our knowledge, the largest collection of choreography models accessible for research purposes.

\subsection{Pre-Processing}
The SAP-SAM dataset contains many non-standard compliant choreography models. For our purposes, we can relax requirements of ownership and observability present in the standard, as the smart contract provides global ownership and observability (c.f.~\cite{ladleifModelingEnforcingBlockchainBased2019}).\footnote{During our initial exploration of the dataset we encountered this issue in many models. The problem is exacerbated by the fact that many participants share the same name in the task bands, but are assigned different participant IDs. We hypothesise that this occurred when a participant's name was entered manually rather than selected from a list of existing ones in the visual editor.\label{fn:many_participants}} 
Furthermore, the models lack execution-relevant information for exclusive gateways (conditions or labels from which conditions could be inferred, and default flow markings). Thus, we pre-processed each model. When no default flow was marked, we set the first outgoing flow to the default flow. Then, for all other outgoing flows, we inserted a boolean condition. Furthermore, we removed any Signavio extension elements. To reduce the size of LLM input, we also removed the BPMN 2.0 Diagram Interchange, as it only contains additional information required to visualise the model. Finally, we merged all start and end events, so each model contains only one, to adhere to the implementation limitation of Chorpiler.

From these pre-processed models, we use Chorpiler to assess the syntactic soundness of each model. Simply put, if Chorpiler is able to generate a contract from the model, we consider it for our benchmark. Chorpiler supports all basic elements of BPMN Choreographies\footnote{We use the latest alpha version (\url{https://github.com/fstiehle/chorpiler/tree/release/v2}, accessed 2025-06-12).} Choreography tasks, start and end event, exclusive, event, and parallel gateways, sub choreographies, and loops in the model. It also ignores issues regarding ownership and observability.

After the filtering steps, \num{1427} choreography models remain.
We use a sample of 165 models for our benchmark runs. 
In our sample, on average, each process contains $13$ participants (see footnote~\ref{fn:many_participants}), six tasks, one diverging exclusive gateway, $0.1$ diverging event-based gateways, and $0.2$ diverging parallel gateways, among others. The largest model in the dataset contains $24$ tasks and ten gateways, respectively.

\subsection{Benchmark}
\label{sec:benchmark}
\paragraph{Prompt.} For our benchmark, we developed multiple prompts. To reduce the likelihood of our results being tainted by a poorly performing prompt, we tested, compared, and refined multiple versions in pre-runs, which we conducted with sets of five to twenty process models. This allowed us to iteratively refine our prompts and test our framework. For any pre-run, in addition to the automated tests, we manually investigated the generated output. This led to many refinements. 
In summary, our prompts moved from loosely defined (zero-shot) instructions, to (better performing) more specific instructions on how the process model should be interpreted and the contract generated. Through our initial tests, we arrived at a one-shot prompt. Specifically, in our prompt we ask for a Solidity implementation for the given process model, enforcing: (i) the control flow, i.e., the order of tasks, (ii) that only the respective initiator can execute a task, and (iii) the autonomous enforcement of gateways, and the evaluation of data-based decisions.\footnote{Our prompt also specifies that the state of the contract should be encoded using a bitmasking technique, as it is the most efficient encoding for a token-based execution (c.f.~\cite{garcia-banuelosOptimizedExecutionBusiness2017a}). This variant did, on average, not perform worse than a prompt asking for a more na\"ive implementation during our pre-runs.} To gauge the effect of prompt-based training, we test a one-shot and a two-shot variant.

\begin{figure}[tb]
    \centering
    \includegraphics[width=1\linewidth]{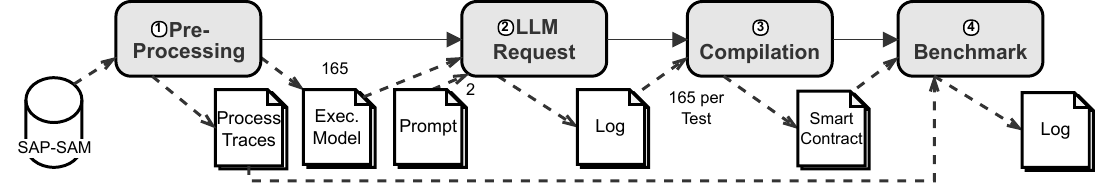}
    \caption{Process of our benchmark experiment: We pre-process the raw model data from the SAP-SAM dataset to receive executable process models and corresponding traces. We use 165 models and perform requests for one and two-shot prompts, a pair for each LLM. From this, logs are generated including the received output. All output is compiled and deployed and benchmarked against the process traces. The benchmark result is logged.}
    \label{fig:process}
\end{figure}

\paragraph{Setup.}
We select a range of top proprietary LLMs\footnote{Our initial run also included \textit{Google's Gemini}. However, we were not able to deactivate the output of its reasoning process, which prevented us from reliably parsing a contract from the output automatically.} as benchmarks and compare them to a host of open source models to evaluate to what extent hosting open source models signifies sacrificing performance for autonomy. Table~\ref{tab:llms} provides an overview of our benchmarked models.
%
\begin{table}[tb]
\centering
\scriptsize
\begin{tabular}{l@{\hskip 1em}l@{\hskip 1em}lr}
\toprule
 & Provider & Model & Size (B) \\
\midrule
Open Source   & DeepSeek  & DeepSeek-V3 0324                  & 671 \\
              & Meta      & Llama-3.1-405b-instruct           & 405 \\
              &           & Llama-3.3-70b-instruct            & 70  \\
              & Alibaba   & Qwen3-235b-a22b                   & 235 \\
\midrule
Proprietary & OpenAI      & GPT-4.1                             & n/a \\
              & Anthropic & Claude Sonnet 4                   & n/a \\
              & X AI      & Grok 3                            & n/a \\
\bottomrule
\end{tabular}
\caption{Models selected for our experiments, conducted June 11--13, 2025, using the then-current models available through OpenRouter. \textit{Size} refers to the number of parameters in billions (B), where available.}
\label{tab:llms}
\end{table}
Figure~\ref{fig:process} gives an overview of our benchmarking experiment run. 
For our LLM selection (c.f. Table~\ref{tab:llms}), we benchmark a one-shot and two-shot variant of our prompt with temperature set to 0,\footnote{Leading to quasi-deterministic (c.f.~\cite{ouyang2025empirical}) inference results. Tie breaking between tokens of equal probabilities, floating point variability, etc. may still cause stochasticity.} on a sample of 165 process models from our pre-processed files.
For the generation of conforming process traces, we set a threshold of \num{2500} traces per process. We generated and replayed $50$ non-conforming traces per process.\footnote{For process models with loops, the threshold is a necessary upper bound. Our non-conforming trace generation algorithm currently depends on a ground truth to assess whether a manipulated trace is not conforming on accident. For models, where the number of conforming traces exceeds our search threshold, non-conforming traces that are actually conforming can be generated. These must be manually removed.
}

We ran our experiment from June 11 to June 13, 2025. For some requests, we had to perform repeated tries, as the provider connection sometimes timed out. We also encountered a period in which OpenRouter experienced an outage. Our framework is set up to retry failed requests on additional runs.

\subsection{Results}
To assess the quality of the produced output we compare usage, in terms of cost and tokens used, as charged by OpenRouter.\footnote{OpenRouter operates on a 'credits' system; credits are purchased upfront, which are then deducted per model request, according to the underlying provider’s token-based rate. OpenRouter charges a fee ($\approx5\% +\$0.35$) when loading credits., which we did not include in our calculations (\url{https://openrouter.ai/docs/faq}, accessed 2025-06-13).} To assess the efficiency of the generated output, we also record gas usage.
%
For a given process model and its generated implementation, we classify the outcome of a trace replay accordingly.
\begin{itemize}
    \item \textit{True Positive}: Each event in a conforming trace was accepted (led to a state change in the contract), and the whole trace led to the end event.
    \item \textit{False Positive}: A non-conforming trace was accepted as per above.
    \item \textit{True Negative}: Any event in the non-conforming trace was rejected, or the trace did not lead to the end event.
    \item \textit{False Negative}: A conforming trace was rejected as per above.
\end{itemize}
%

Using this classification framework, we calculate precision and recall per process case, using standard formulations,  and the F1 score (the harmonic mean of precision and recall), common metrics to assess LLM output (see e.g.,~\cite{chang2024survey}). In our case, F1 is a suitable choice over metrics like accuracy, since the number of traces is unbalanced.

To assess the overall performance of each model, we calculate the macro F1 (the average of all F1 across all cases). 
Our results are shown in Table~\ref{tab:result}. The most obvious aspect is that Grok and Claude achieved F1 scores of 0.8 or more in all variants. Most of the generated code compiled. The cost for translating a process model was on the orders of 0.1 cents to a few cents. Interestingly, the two-shot prompt did not consistently yield better results.
As a side effect of the experiment, we observed that the framework performed well in running the benchmark across the diverse set of process and AI models. 
\begin{table}[tb]
\centering
\scriptsize
\renewcommand{\arraystretch}{1.2}
\begin{tabular}{l @{\hskip 1em} l@{\hskip 2em} r@{\hskip 1em}r @{\hskip 2em} rr}
\toprule
\textbf{Shot} & \textbf{Model} & \multicolumn{2}{l}{\textbf{(Avg./Process)}} & \multicolumn{2}{c}{\textbf{Correctness}} \\
\cmidrule{3-4} \cmidrule{5-6}
& & Cost(\$) & Tokens & F1 Mac. & Comp.(\%) \\
\midrule
One & grok-3-beta             & \num{0.044} & \num{10,134}   & \num{0.918} & \num{100.0} \\
    & claude-sonnet-4         & \num{0.046} & \num{11,442}     & \num{0.862} & \num{100.0} \\
    & gpt-4.1                 & \num{0.028} & \num{10,326}   & \num{0.797}  & \num{99.4} \\
    & qwen3-235b-a22b         & \num{0.009} & \num{18,444}     & \num{0.648} & \num{97.0} \\
    & deepseek-chat-v3-0324   & \num{0.005} & \num{10,483}     & \num{0.580} & \num{99.4} \\
    & llama-3.1-405b-instruct & \num{0.010} & \num{10,259}      & \num{0.475 } & \num{99.4} \\
    & llama-3.3-70b-instruct  & \num{0.001} & \num{10,249}    & \num{0.399} & \num{90.4}  \\
\midrule
Two & grok-3-beta             & \num{0.056} & \num{14,964}  & \num{0.861} & \num{100.0} \\
    & claude-sonnet-4         & \num{0.064} & \num{17,218}    & \num{0.853} & \num{100.0} \\
    & gpt-4.1                 & \num{0.038} & \num{15,410}   & \num{0.696} & \num{100.0} \\
    & deepseek-chat-v3-0324   & \num{0.007} & \num{15,680}  & \num{0.669} & \num{97.6} \\
    & qwen3-235b-a22b         & \num{0.009} & \num{24,415}   & \num{0.581} & \num{93.4} \\
    & llama-3.1-405b-instruct & \num{0.016} & \num{15,473}     & \num{0.431} & \num{98.8} \\
    & llama-3.3-70b-instruct  & \num{0.002} & \num{15,252}  & \num{0.370} & \num{97.0} \\
\bottomrule
\end{tabular}
\caption{Result of our benchmark run with $165$ process models. We report the average cost in US-\$ (as reported by OpenRouter), and tokens used per process model. The overall correctness is reported via the F1 macro. Compilability (Comp.) reports on the percentage of syntactically correct generated contracts.}
\label{tab:result}
\vspace{-1em}
\end{table}

%
%
%

%
\vspaceSecBefore
\section{Discussion, Limitations \& Future Work}\label{sec:discussion}
\label{sec:discussion}
\vspaceSecAfter
%



Our results show that current LLMs 
can transform executable choreography models into syntactically and functionally correct 
smart contracts most of the time---even when benchmarked against a realistic and diverse dataset. 
While the results in terms of F1 score shows promise, it can only serve as a point of departure for future work exploring meaningful integrations within a smart contract generation workflow. 
As mentioned earlier, blockchain is an unforgiving environment, and smart contract vulnerabilities may become very costly. 
Indeed, the goal of business process execution via smart contracts is to provide a trusted and secure decentralised platform. These requirements would be undermined by a reckless integration of LLM capabilities.

F1 scores that do not reliably achieve 100\% would not be suitable for this context.
Say, the approaches would be improved to achieving an average F1 score of 98\%; while this would be impressive in many domains, it falls short of the perfect reliability required for blockchain-based smart contracts. Given the financial risks and immutable nature of blockchain transactions, even such a 2\% error rate could lead to significant vulnerabilities or losses, making such performance inadequate for real-world deployment.
We do not see a way in which this fundamental issue could be resolved with current LLM architectures.

However, we see ways to advance the current direction beyond using output as-is.
Future work should explore integrating LLM capabilities into existing smart contract generation tools.
For smart contract development, which demands high security, LLM integration must rely on extensive evaluation and robust verification of generated outcomes. This could involve using LLMs to propose code snippets or modifications, which are then rigorously checked against formal specifications (as we demonstrated with our framework) or verified using automated theorem provers, before being considered. LLMs should also generate test cases or identify potential vulnerabilities themselves, identifying common or context-depending issues in smart contracts, augmenting existing verification processes.




Furthermore, LLMs could be used to extend the functionality of existing code generation tools (generating new rules) by: (i) generating more flexible templates based on specific process models, (ii) generating code snippets for edge cases not covered by standard templates, (iii) suggesting optimisations for rules and generated code, and (iv) assisting in translating between different smart contract languages or blockchain platforms. Any such use would still have to be vetted in a suitable form, but has the advantage that rule improvements and extensions are not subject to non-determinism after being included in the code generation tools. This approach combines the reliability and domain-specific knowledge of traditional code generation tools with the flexibility and natural language understanding of LLMs, and hence addresses the limitations mentioned in the introduction. 

Our benchmarking framework can serve as a foundation for evaluating these future directions, helping to assess the quality of LLM-generated code, the effectiveness of verification methods, and benchmarking extended code generation tools. Towards this, the capabilities of the benchmarking tool itself must be extended to, e.g., consider other factors such as efficiency of generated code, potential biases present in the LLM, or sustainability factors (c.f.~\cite{haase2025sustainability}).

Finally, some limitations apply to our performed benchmark. We experimented with different prompts, but cannot guarantee that the selected query was optimal; furthermore, we used the same prompt across all LLMs, which may be suboptimal. Although we included a diverse set of LLMs, our findings should not be assumed to generalise to all current or future LLMs.




%
\vspaceSecBefore
\section{Conclusions}\label{sec:concl}
\vspaceSecAfter
In this work, we presented an exploratory study investigating the use of LLMs for generating smart contract code from business process descriptions. We introduced an automated evaluation framework and provided empirical data from a large dataset of 165 process models. Our results show that while current LLMs can transform executable choreography models into syntactically and functionally correct smart contracts most of the time, achieving F1 scores of 0.8 or more for top-performing models, this performance falls short of the perfect reliability required for smart contracts. Given the financial risks and immutable nature of blockchain transactions, even small error rates could lead to significant vulnerabilities or losses. We argue that this fundamental issue cannot be resolved with current LLM architectures. Instead, we propose future work to explore responsible LLM integrations in existing tools for code generation, focusing on using LLMs for verification and enhancing current code generation tools rather than replacing them entirely. Our benchmarking framework can serve
as a foundation for developing and evaluating such integrations.
%
\vspace{-.6em}
\credits{
\subsubsection*{\ackname}
Generative AI was used to assist in the editing of the manuscript and the implementation of the artifact.
Generated output was never taken "as-is", it was reviewed and verified by the authors.}
%
%
%
\bibliographystyle{splncs04}
\bibliography{bib}
\end{document}